\documentclass{aa}  
\def\code#1{\texttt{#1}}
\usepackage{natbib}
\makeatletter
\newcommand*{\rom}[1]{\expandafter\@slowromancap\romannumeral #1@}
\usepackage[version=4,arrows=pgf-filled]{mhchem}
\makeatother
\DeclareUnicodeCharacter{2212}{-}
\bibpunct{(}{)}{;}{a}{}{,} 
\usepackage{graphicx}
\usepackage{txfonts}
\usepackage{siunitx}
\usepackage{xcolor}

\begin{document}

   \title{Investigating the complex absorbers of Mrk\,766 with XMM-\textit{Newton}}

   \author{T. Matamoro Zatarain
          \inst{1,2,3}
          \and
          E. Costantini 
          \inst{1,2}
          \and
          A.  Jur\'{a}\v{n}ov\'{a}
          \inst{4,1,2}
          \and
          D. Rogantini
          \inst{5}
          }

   \institute{SRON Netherlands Institute for Space Research, Niels Bohrweg 4, 2333 CA Leiden, The Netherlands
         \and
             Anton Pannekoek Institute, University of Amsterdam, Postbus 94249, 1090 GE Amsterdam, The Netherlands
        \and School of Physics, HH Wills Physics Laboratory, University of Bristol, Tyndall Avenue, Bristol, BS8 1TL, UK
        \and MIT Kavli Institute for Astrophysics and Space Research, Massachusetts Institute of Technology, Cambridge, MA 02139, USA
        \and Department of Astronomy and Astrophysics, University of Chicago, Chicago, IL 60637, USA    
             }

   \date{Received}

 
  \abstract
   {}
   {We examine the high energy resolution X-ray spectrum of the narrow-line Seyfert 1 galaxy Mrk\,766 using 4 observations taken with \rm{XMM}-\textit{Newton} in 2005, to investigate the properties of the complex ionised absorber / emitter along the line of sight, as well as absorption by dust intrinsic to the source. }
   {We make use of the high-energy resolution RGS spectrum to infer the properties of the intervening matter. We also use the spectrum obtained by EPIC-pn and the photometric measurements of OM to obtain the spectral energy distribution of the source, necessary for the photoionisation modelling of the ionised outflow.}
   {The warm absorber in Mrk\,766 consists of two phases of
photoionisation. In addition to these two warm absorber components with $\log\xi\sim 2.15$ and $\log\xi\sim -0.58$, we find evidence of absorption by a collisionally ionised component ($T\sim51$ eV). We discuss the implication of this additional component in light of theoretical predictions. Moreover, we detect signs of absorption by a dusty medium with $N_\text{dust}\sim 7.29\times 10^{16}$ cm$^{-2}$. Finally the relatively weak emission features in the spectrum seem to be unrelated to the absorbers and probably originated by an out-of sight-line ionised plasma.}
{}


   \maketitle
%

\section{Introduction}

It is standard for Seyfert galaxies' spectra to show absorption features due to photoionised gas in our line of sight \citep{Kaastra_2000, Blustin_2005, Constantini_2007, Mehdipour_2010}. This outflowing gas, often referred to as warm absorbers (WAs), is particularly common in Seyfert 1 galaxies \citep{Reynolds_1995, Laha_2014, Laha_2020}. WAs imprint narrow absorption lines and edges on the X-ray spectrum of active galactic nuclei (AGN). In most cases, more than one phase of WA is present in the AGN, where each phase constitutes a region of the gas that is outflowing with a specific ionisation parameter ($\xi$), column density ($N_\text{H}$), and velocity ($v_{\text{out}}$).

AGN outflows are thought to provide a link between the co-evolution of supermassive black holes (SMBHs) and their host galaxies \citep{Fabian_2012, King_2015}. They are suggested to be responsible for regulating black hole growth \citep{Crenshaw_2012} and for helping establish the $M-\sigma$ relation \citep{Silk_1998, King_2015}. Moreover, they may also be responsible for the quenching and / or enhancement of star formation, as they are able to enrich the gas in the galaxy and / or remove gas from the bulge and halo \citep{Hopkins_2005, Heckman_2014}. 

The study of outflows is therefore important to establish this connection between SMBH and host galaxy. In particular, WAs provide us with insights into some of the innermost regions of AGN. However, there are still many open questions regarding the nature of WAs, including their origin, their geometrical structure, or their connection to the nucleus and outer regions of the AGN. Some past studies connect the origin of these absorbers to photoionised evaporation from the inner edge of the torus \citep{Blustin_2002, Mckernan_2007}. Meanwhile, other works have attempted to explain their geometry in the context of ionisation cones \citep{Kinkhabwala_2002}, whereby depending on the inclination of the AGN it is possible for the WAs to appear in re-emission \citep{Kinkhabwala_2002, Bianchi_2006,Guainazzi_2007}.

Nevertheless, not only photoionisation is responsible for the ionised absorbing gas detected in the X-ray spectra of AGN. In fact, various AGN have shown \ion{Fe}{}\ absorption features that have been associated to the presence of dust \citep{Lee_2001, Lee_2013, Mehdipour_2018, Psaradaki_2024}. Moreover, signatures of collisionally ionised absorbers (CIAs) have been recently reported in NGC 4051 \citep{Ogorzalek_2022}. Other sources presenting collisionally ionised absorbers include NGC 4151 \citep{Wang_2011}, NGC 3393 \citep{Maksym_2019}, and Mrk 573 \citep{Paggi_2012}. The distance, and therefore origin, of these collisionally ionised absorbers are extremely difficult to derive because of our current instrumental resolution, making it hard to learn more about their energetics and geometrical properties.

There is also evidence that AGN, mostly Seyfert 1s, can host narrow emission lines of \ion{C}\ and \ion{O}\ embedded in the absorption features produced by the WAs \citep{Bianchi_2006, Guainazzi_2007, Ebrero_2010}. Whether these lines are connected to the WAs or not is still an open question \citep{Costantini_2010}. However the emission spectrum, and its link to the WA, could help us to obtain a better understanding of the sort of processes that ionise these outflows, and the distances at which they originate. \cite{Guainazzi_2007} conducted a study on obscured AGN, reporting the presence of emission in the form of radiative recombination continua (RRCs), which suggest that these features are likely created by photoionised gas. They demonstrated that resonant scattering significantly influenced the ionization / excitation balance of the gas. Their results indicated that the column densities required for such emission matched those measured in WAs observed in unobscured AGN. However, the observed emission in AGN might originate from different regions within the system.

Depending on the type of AGN, it is possible that the appearance of these emission lines will differ. In the case of highly obscured Seyfert 2 AGN, their soft X-ray spectrum is dominated by the emission from the \ion{H}-like, \ion{He}-like K transitions of light metals, likely created by photoionised gas surrounding the AGN around the NLR \citep{Bianchi_2006}. This  is due to the torus obscuring all evidence of absorption affecting the continuum. However, in the case of Seyfert 1 AGN the continuum emission is predominant as the torus is not obscuring our line-of-sight, implying favourable detection of absorption instead of emission \citep{Costantini_2010}. As the inclination of a source increases to somewhere between a Seyfert 1 and Seyfert 2, it is possible, to an extent, to disentangle those emission lines from the NLR from the different absorption features affecting the continuum.

In this work, we explore the presence of the complex absorbers in the soft X-ray spectrum of Mrk\,766. This AGN is a nearby narrow-line Seyfert 1 galaxy \citep[$z \sim 0.0129$,][]{Falco_1999} with a SMBH of $M_\text{BH}\sim1.26\times10^{6}M_{\odot}$ hosted in a barred spiral galaxy \citep{Risaliti_2010, Giacche_2015}. It is one of the brightest sources of its class, with $F_{2-10\text{keV}}\sim1.37\times 10^{-11}$ erg cm$^{-2}$ s $^{-1}$ (this work), and has been observed by multiple X-ray observatories in different occasions. Previous studies have shown that the X-ray spectrum of Mrk\,766 is variable and that it can be modelled with an absorbed power law and an accompanying ionised reflection \citep{Turner_2007}. Moreover, the presence of WAs in its X-ray spectrum has long been established \citep{George_1998, Mckernan_2007, Laha_2014, Winter_2012}, although past \rm{XMM}-\textit{Newton} \citep{Jansen_2001} observations caused a debate on whether the complicated features in the soft X-ray spectrum of this source arise from a dusty warm absorber or relativistic emission lines \citep{Raymont_2001, Lee_2001, Sako_2003, Mason_2003}.

Throughout this work, we adopt the following cosmological parameters: $H_0=70$ kms$^{-1}$ Mpc$^{-1}$, $\Omega_m =0.3$, and $\Omega_\Delta =0.7$. The analysis was carried out using the fitting package SPEX v3.08.00 \citep{Kaastra_1996, Kaastra_2017}. For evaluating the goodness of fit, we used the \textit{C}-statistic \citep{Cash_1979, Kaastra_2017} and the reported uncertainties are calculated with $1\sigma$ significance.


\section{Observations and data processing}

\begin{figure}
	\includegraphics[scale=0.25]{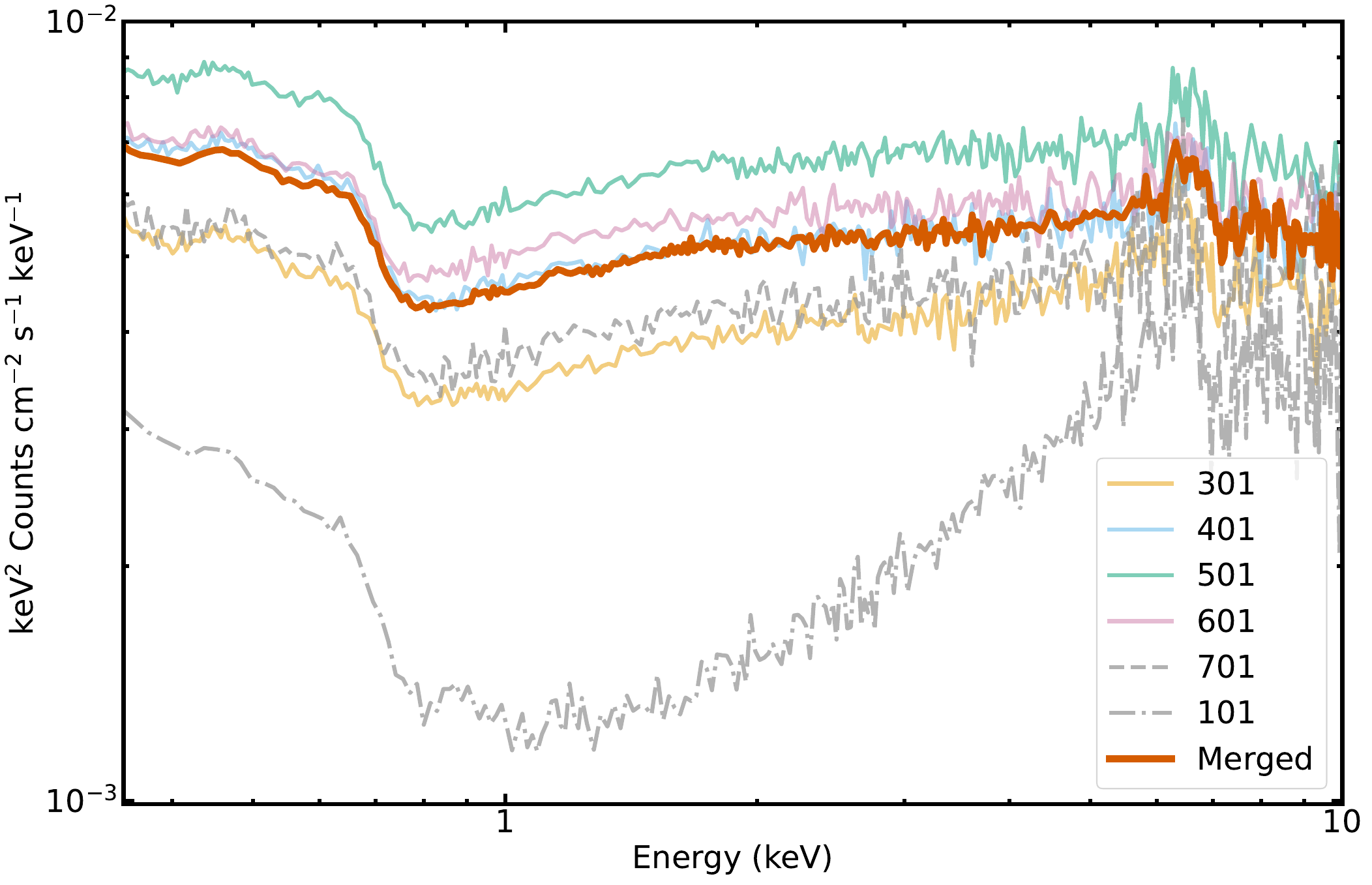}
    \caption{Spectra of Mrk\,766 taken with the \rm{XMM}-\textit{Newton} EPIC-pn instrument. The different observations taken during the 2005 campaign are shown in various colours. In grey we show those observations that were not included in the study. In red we portray the merged spectrum of observations $301, 401, 501$ and $601$. We refer to the observations by the last three digits of their observation IDs. }
   \label{epic-observations}
\end{figure}

Mrk\,766 was observed by \rm{XMM}-\textit{Newton} over a period of $6$ orbits between May 23$^{\text{rd}}$ 2005 and June 3$^{\text{rd}}$ 2005, of which we chose to analyse $4$ observations which had the longest exposure times, comparable fluxes, and spectral shapes. We did not choose to use all $6$ observations taken during this \rm{XMM}-\textit{Newton} campaign since we were interested in exploring the average properties of the AGN and its absorbers. Therefore, we decided to exclude observation $0304030101$ due to its notably lower flux and different spectral shape (see Fig. \ref{epic-observations}) as well as observation $0304030701$ because of its particularly shorter exposure-time ($\sim 1/3$ of the average exposure time). The details for the chosen observations with their corresponding observation IDs, observation dates, and exposure times can be found in Table \ref{observations}. We used data from EPIC-pn  \citep{Turner_2001} in the $0.3-10$ keV band, as well as data from the Reflection Grating Spectrometer (RGS) \citep{Herder_2001} in the $7-36$ \AA\ band, and the Optical Monitor (OM) \citep{Mason_2001} filters B, U, UVW1, UVM2, and UVW2. We reduced the data using the Science Analysis System (SAS ver $20.0.0$) and the HEASARC FTOOLS.  

The EPIC-pn camera was operated in small window mode. We verified using the SAS task \code{epatplot} that the observations were not affected by pile-up. Based on a $3\sigma$ clipping of the background count rate, the data from neither observation was significantly affected by flares. We extracted the spectrum from a circular region, centred around the source, of 30.6 arcsec. We used a region of the same radius to extract the background from a source-free area on the same chip. We created out-of-time event lists using the \code{epchain} task and, after filtering, we rescaled and removed them from the good-time intervals. Finally, we created the ancillary response file (ARF) and the redistribution matrix file (RMF) using the SAS commands \code{arfgen} and \code{rmfgen} respectively. 

The RGS instruments were operated in the Spectroscopy HER + SES mode, and the data were processed with the \code{rgsproc} task. We created good-time intervals with the \code{tabgtigen} command, and generated light curves for RGS1 and RGS2 with \code{eveselect} and corrected them with \code{rgslccorr} for dead time, background scale, whilst performing background subtraction. Finally, we combined RGS1 and RGS2 with \code{rgscombine}.

The OM observations  were obtained using the Imaging and Fast modes. Both of these modes were used together for most of the observations, where the filters B, U, UVW1,UVM2, and UVW2 were applied, except for observation  $0304030301$ which only utilised the Image mode with the grism filters and, therefore, could not be used for the purpose of this analysis. We used the SAS task \code{omichain} to filter the source lists.  We combined the source list files with the coordinates for the source, which were cross-checked using the SIMBAD astronomical data base\footnote{\url{http://simbad.u-strasbg.fr/simbad/sim-basic?Ident=Mrk\,766&submit=SIMBAD+search}.}, to obtain the raw count rates and the errors using the task \code{om2pha}.

\begin{table}
\caption{\rm{XMM}-\textit{Newton} observations used in this work.}\label{observations}

\centering                                   

\begin{tabular}{c c c}         
\hline\hline                       

Observation ID & Observation date  & Net exposure time \\
&(yyyy-mm-dd) &(s)\\

\hline            

0304030301 & 2005-05-25 & 62228 \\
 
0304030401 & 2005-05-27 & 61281 \\

0304030501 & 2005-05-29 &  62914 \\

0304030601 & 2005-05-31 &  50814 \\
\hline                                             
\end{tabular}
\end{table}

 We merged the different EPIC-pn, RGS, and OM observations respectively using SAS an created a combined set of response and spectrum files for each instrument. By doing so we obtained a better signal-to-noise ratio to work with, which allowed us to be able to find the average characteristics of the present absorbers. The individual observations and merged spectrum can be seen in Fig. \ref{epic-observations}.

\section{Spectral energy distribution}\label{Sect 3}

\begin{figure}
	\includegraphics[scale=0.25]{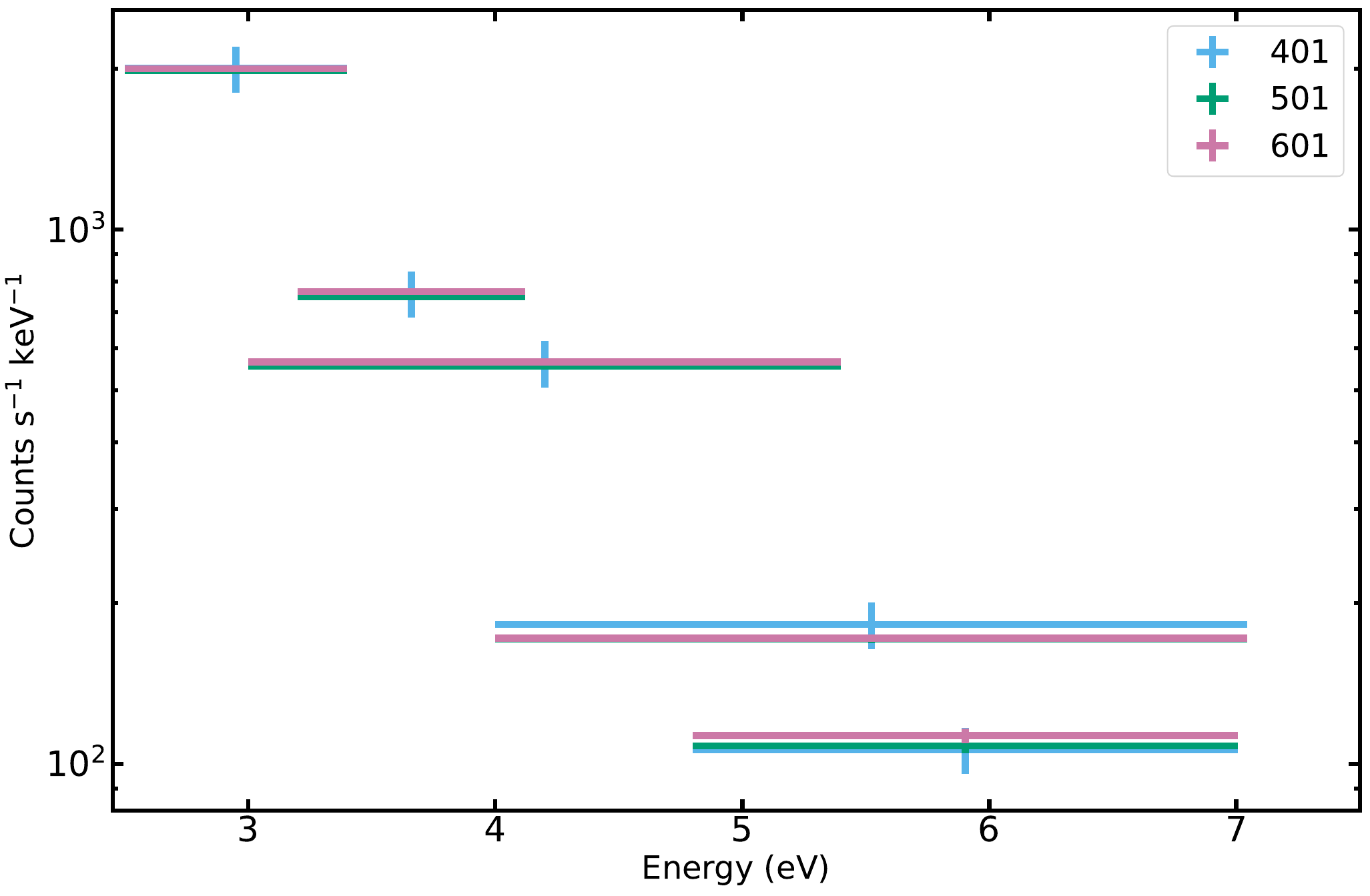}
    \caption{Spectra of Mrk\,766 taken with the \rm{XMM}-\textit{Newton} OM instrument. The different observations used for this study are shown in various colours. The OM filters from left to right are B, U, UVW1, UVM2, UVW2. We refer to the observations by the last three digits of their observation IDs.}
   \label{OM_observations}
\end{figure}

For the spectral energy distribution (SED), we used the merged EPIC-pn observations. As the OM photometric data did not change across the observations, we adopted the data from the 0304030401 observation (see Fig. \ref{OM_observations}). For modelling, we followed the approach presented in e.g. \citet{mehdipour_2011}, where the broad band continuum is described by a seed black body from the accretion disk \citep[{\sc dbb} model in SPEX,][]{SS_1973}. This radiation is then reprocessed by inverse Compton scattering \citep{Done_2012, Kubota_2018}, creating a warm soft-X-ray emission component ({\sc comt} model in SPEX). The X-ray continuum has been parametrized by a cut-off powerlaw and a reflection component \citep[{\sc refl} model in SPEX,][]{Zycki_1999}.

We use the information obtained from the RGS fitting for the complex absorption components, needed to obtain the best broad band fit. We fixed to the RGS best fit values the Galactic cold and warm absorption, and the dust absorbers in the host galaxy (Sect.~\ref{RGS}, Table~\ref{table:4}). At the end of this iterative process, the parameters of the warm absorbers detected by  RGS will be refined by the ionization balance dictated by the broad band SED. We added two photoionised absorbers to model the different phases of WAs and a collisionally ionised absorber to account for the additional absorption in the RGS data (Sect.~\ref{RGS}) leaving the column density, the ionization parameter of the WAs, and the temperature of the CIA free to vary, starting from the best fit values. The free parameters adjust to the different resolution of EPIC-pn and any cross calibration uncertainties that may still exist. Finally, for the OM data, we included an extinction component to account for the local extinction \citep[$E(B-V)$=0.017][]{Schlafly_2011}, assuming $R_V=3.1$ and Milky Way extinction law \citep{Cardelli_1989}. A larger extinction \citep[$E(B-V)\sim0.38$][]{Vasudevan_2009} component at the redshift of the source was also added. In this case a SMC extinction law \citep{Gordon_2003}, believed to be more descriptive of extinction in AGN \citep[][]{Hopkins_2004}, was adopted. The OM filter B was excluded from the analysis as the flux could be still contaminated by the star radiation in the host galaxy. 

The broad-band continuum best fit was obtained with a disk black body with a temperature $T_{\rm dbb}=10.4\pm0.2$\,eV. This value was taken as the seed temperature for the Comptonisation component, with an optical depth $\tau = 10.4\pm0.1$ and a final temperature of $T=0.34\pm0.02$\,keV. The high energy cut off of the powerlaw, with slope with $\Gamma=1.90\pm0.02$, remains unconstrained. We fixed the cut off to 22\,keV, keeping in mind that a large uncertainty exists on this value \citep{Buisson_2018}. Finally, the fit required a modest reflection component, with an associated Fe\,K$\alpha$ emission line. The inclination of the disk was set at $46^{\circ}$  \citep{Buisson_2018}, while the scale parameter in the {\sc refl} model was $s<3\times10^{-3}$ (for reference, $s=1$ would correspond to an isotropic source above the disk). Other parameters were kept at default values, as the SED modelling is limited by the lack of hard X-ray data, therefore making it hard to determine the reflected spectrum of the source. An additional narrow Gaussian profile was required at $E\sim6.7$\,keV. 

\begin{figure}
	\includegraphics[scale=0.37]{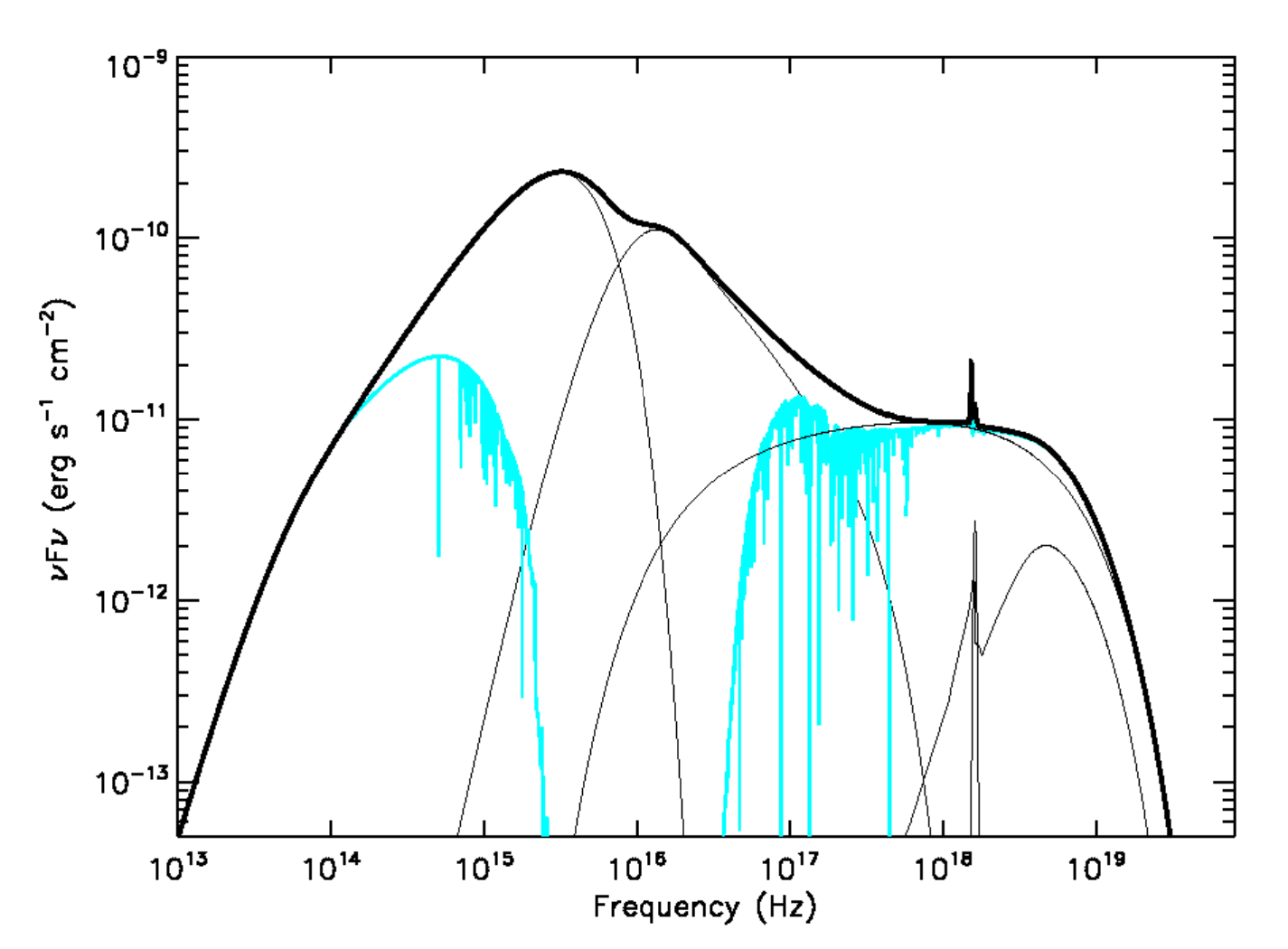}
    \caption{Model obtained from the SED fitting of the \rm{XMM}-\textit{Newton} EPIC-pn and OM merged observations. In black we show the best-fit model to the data. The curves in grey showcase the emission components that make up the model, which from left to right are {\sc dbb}, {\sc comt}, {\sc pow}, {\sc refl}, and {\sc gauss}. Blue represents the model of the absorbed spectrum.}
   \label{SED}
\end{figure}

The resulting SED is presented in Fig. \ref{SED}. Its ionising luminosity, defined as the luminosity between $1$ and $1000$ Ryd, is $L_{\rm ion} = 1.58 \times 10^{44}$ erg s$^{-1}$. This SED was then used with \code{xabsinput} in SPEX to calculate the ionisation balance that accompanies the {\sc xabs} models in Sect. \ref{RGS}.

\section{RGS spectral analysis}\label{RGS}

\begin{figure}
    \centering
    \includegraphics[scale=0.35]{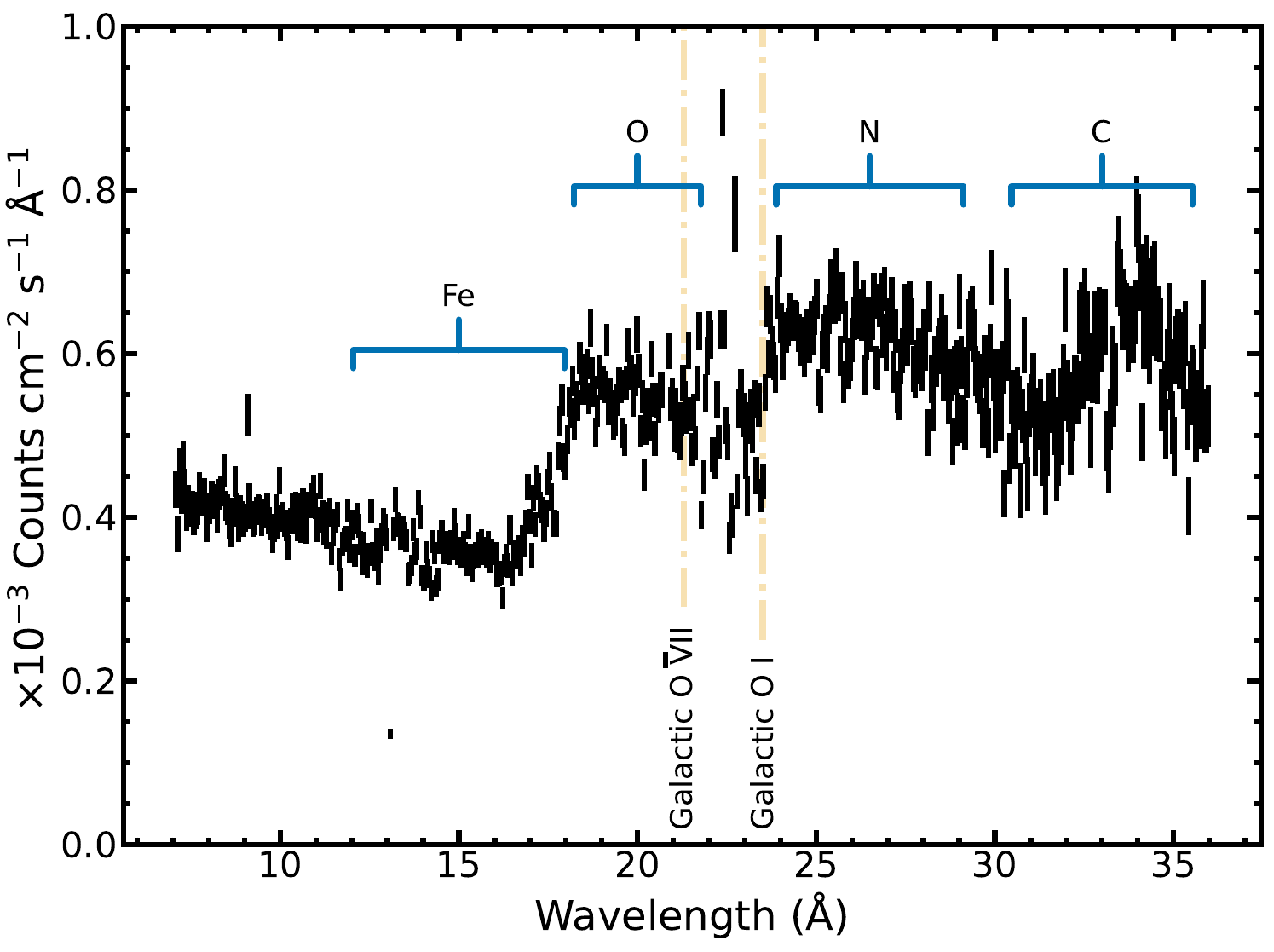}
    \caption{Merged RGS data in black. We indicate with horizontal lines the regions where we expect ions of a certain element to show in the spectrum. Features marked in yellow correspond to Galactic features, whilst blue indicate regions where absorption / emission features by certain elements are expected.} 
    \label{RGS_absorption}
\end{figure}

For simplicity, and given the limited band of RGS, we modelled the continuum with a simple broken power-law model. In this model we left the two slopes free to vary as well the break energy. Given the prominent soft excess in this source, this phenomenological model mimics, in the RGS band, the curvature given by the broader band Comptonisation model (Tab. \ref{table:4}). 

\subsection{Absorption features}

Various types of absorption and emission signatures are present in the data. Fig. \ref{RGS_absorption} presents the merged RGS observations, where numerous spectral features can easily be distinguished by eye. Starting with the $16-23$ \AA\, region, absorption edges are apparent, which are consistent to be produced by Fe and O. This can be taken as a sign of dust being present in the line of sight \citep{Lee_2001,Lee_2013,Mehdipour_2018}. Moreover, the apparent absorption features at the laboratory wavelengths of \ion{O}{i} and \ion{O}{vii} indicate the presence of both neutral and ionised gas intrinsic to the Milky Way \citep{Steenbrugge_2005,Constantini_2007, Mehdipour_2018}. Furthermore, there is a plethora of other type of narrow absorption features that could be linked to the presence H and He-like transitions from C, N, O, \,and Ne, such as \ion{Ne}{X}, \ion{Ne}{ix},\,or \ion{O}{viii} amongst others.

To account for neutral absorption in the Galaxy, we used the {\sc hot} component in SPEX with a temperature fixed at $1\times10^{-6}$ keV. To portray the absorption caused by a warmer ISM gas, and therefore take into account the apparent \ion{O}{vii} absorption feature at wavelength $21.6$\AA, we added a second {\sc hot}. We allowed for the column density of this component to be free, but limited its temperature within the range ($5-7.5$) $\times 10^{-2}$ keV \citep{Yao2006}.

To investigate the presence of WAs, we modelled their absorption features using the {\sc xabs} component, which calculates the transmission of a slab of an ionised material, where all ionic column densities are linked through a photoionisation model \citep{Steenbrugge_2005}. The provided SED dictates the ionisation balance, which sets the fractional abundance of different charge states of a particular element. We allowed the column density ($N_\text{H}$), ionisation parameter ($\log\xi$), outflowing velocity ($v_{\text{out}}$), and covering factor (C$_V$) of the absorbers to be free to vary. We initially fitted a single WA with $\log\xi=0.88\pm0.17$, which provided us with a $C$-stat/d.o.f. of $2081/561$. However, from the residuals it became obvious that additional absorption features were present in the lower and higher energy ranges of the RGS data. Therefore, we introduced two additional WAs with a higher and lower $\log\xi$ values respectively to cover the different energy regions. We found an ionised absorber with $\log\xi= 2.18\pm 0.05$, and another at $\log\xi= 0.92 \pm 0.02$. These two components were found to not fully cover the source, with a measured covering factor of $0.27 \pm0.01$ and $0.70\pm 0.07$ respectively. Finally, for the third lower ionization component ($\log\xi= -1.34\pm 0.25$), the covering factor was consistent with unity. These additional two WAs components improved the $C$-stat/d.o.f. from $2081/561$ to $1564/553$.

\begin{figure}
    \centering
    \includegraphics[scale=0.24]{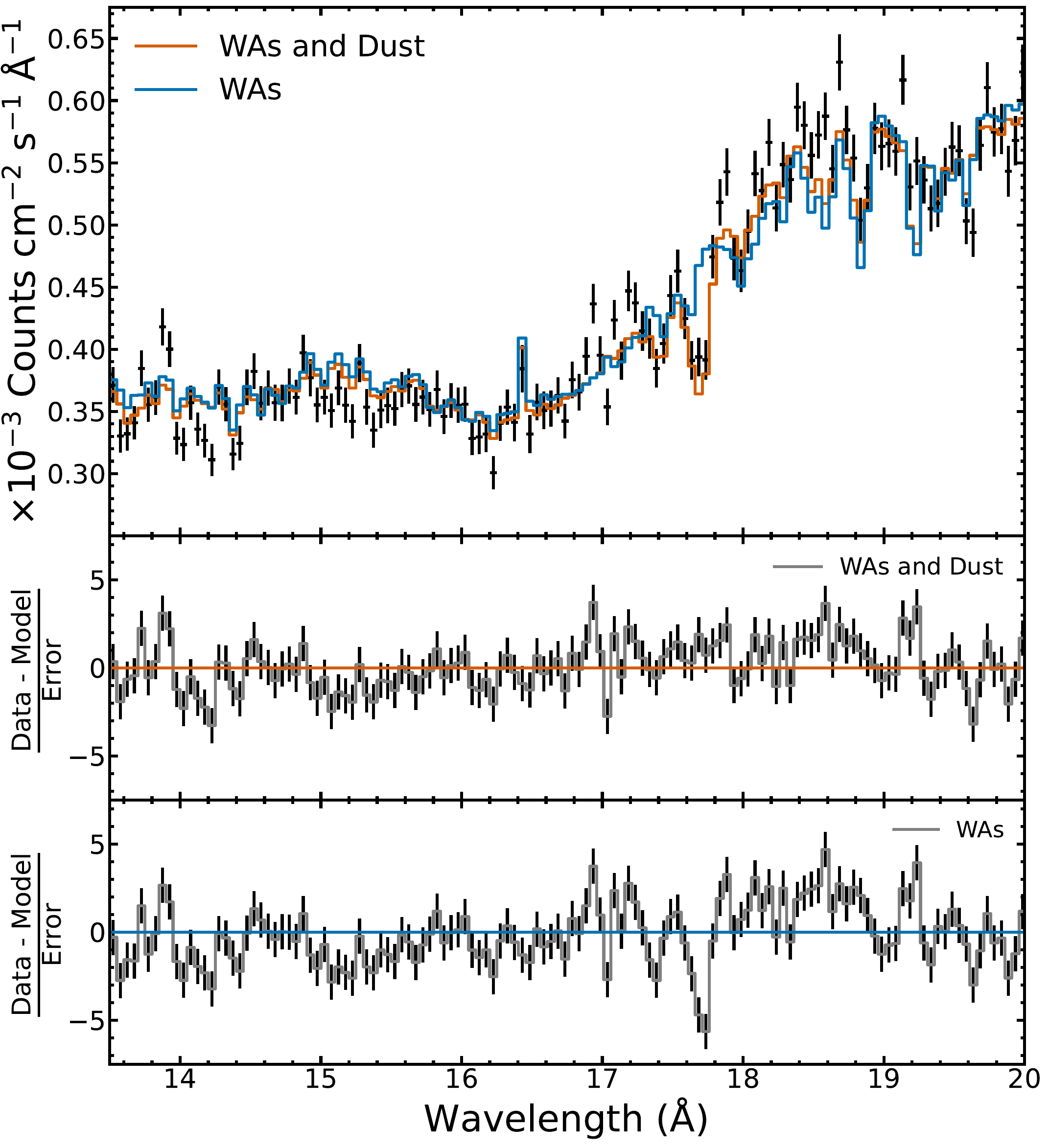}
    \caption{\textbf{Upper panel}: Comparison between a model using a combination of three WAs and dust (red) as opposed to only WAs (blue). It is apparent that the latter case does not take into account the additional absorption in the $15-18$ \AA\ region, while the WAs and dust model does. The RGS data is shown in black. \textbf{Middle panel}: Residuals of the WAs and dust model. \textbf{Lower panel}: Residuals of the WAs model. } 
    \label{dust}
\end{figure}

Although introducing the different phases of WAs did improve the overall fit to the data, the residuals showed that there was still additional absorption present that needed to be considered, particularly in the $16-20$ \AA\ region, as can be seen in the lower panel of Fig. \ref{dust}. To account for the deep absorption jump at $\sim17.7$ \AA, we considered absorption by dust at the redshift of the source, i.e. within the system of Mrk\,766. We chose to include a dust component ({\sc amol} in SPEX) instead of another WA because of the potential \ion{Fe}{i} absorption features present in the spectra. These lines could either be caused by a cold absorber or by dust. Dust features in the RGS band consist mainly of absorption by Fe ($17.71$ \AA) and O ($22.83$ \AA). Therefore, we assumed a hematite composition \citep[\ce{Fe2O3},][]{Mehdipour_2010}. The addition of this component improved our fit from a $C$-stat/d.o.f. of $1564/553$ to $1325/551$, and took into account the deep absorption jump at $\sim17.7$\AA. Moreover, it affected the ionisation values the absorbers, and shifted them to $\log\xi= 2.39\pm 0.05$, $\log\xi= 0.96 \pm 0.02$, and $\log\xi= -0.80\pm 0.15$. Fig. \ref{dust} showcases how the inclusion of a dust component improved the overall fit to the data over this region. 

However, the residuals of such a model seemed to still show an excess in absorption all over the $17-19$ \AA\ region. To mitigate this additional absorption, we tested the presence of a collisionally ionized absorber. In particular, we substituted the WA with $\log\xi\sim 0.96$ by a warm CIA (modelled with the SPEX component {\sc hot}), intrinsic to Mrk\,766. We found that by adding this CIA we improved our results over the $17-29$ \AA\ region. With a temperature $T=0.037\pm0.002$ keV, column density of $N_{\text{H}}= (1.22\pm0.04)\times 10^{22}$ cm$^{-2}$, and a partial covering of C$_V$ $= 0.53\pm0.09$, we found that such a CIA provided a good fit to the data, where a model with only WAs created a deeper absorption over this region. At a given temperature, the ionic column density distribution of a collisionally ionised gas peaks around a single ion, while for a photoionised gas, a range of ions is involved, leading to additional absorption when using only {\sc xabs} \citep{van_Peet_2009}.  In fact, statistically speaking, the inclusion of a collisionally ionised absorber improved our $C$-statistic, with $C$-stat/d.o.f. $=1243/551$, as opposed to $1325/551$ for a model that only uses WAs without CIAs. To ascertain that a model with a CIA was statistically better, we used the Akaike information criterion (AIC), by which if we have a collection of models for a given set of data, we can estimate the quality of each model relative to each other by investigating the relative amount of information lost by a model, where the less information lost the better \citep{Akaike}. 

\begin{figure}
	\includegraphics[scale=0.25]{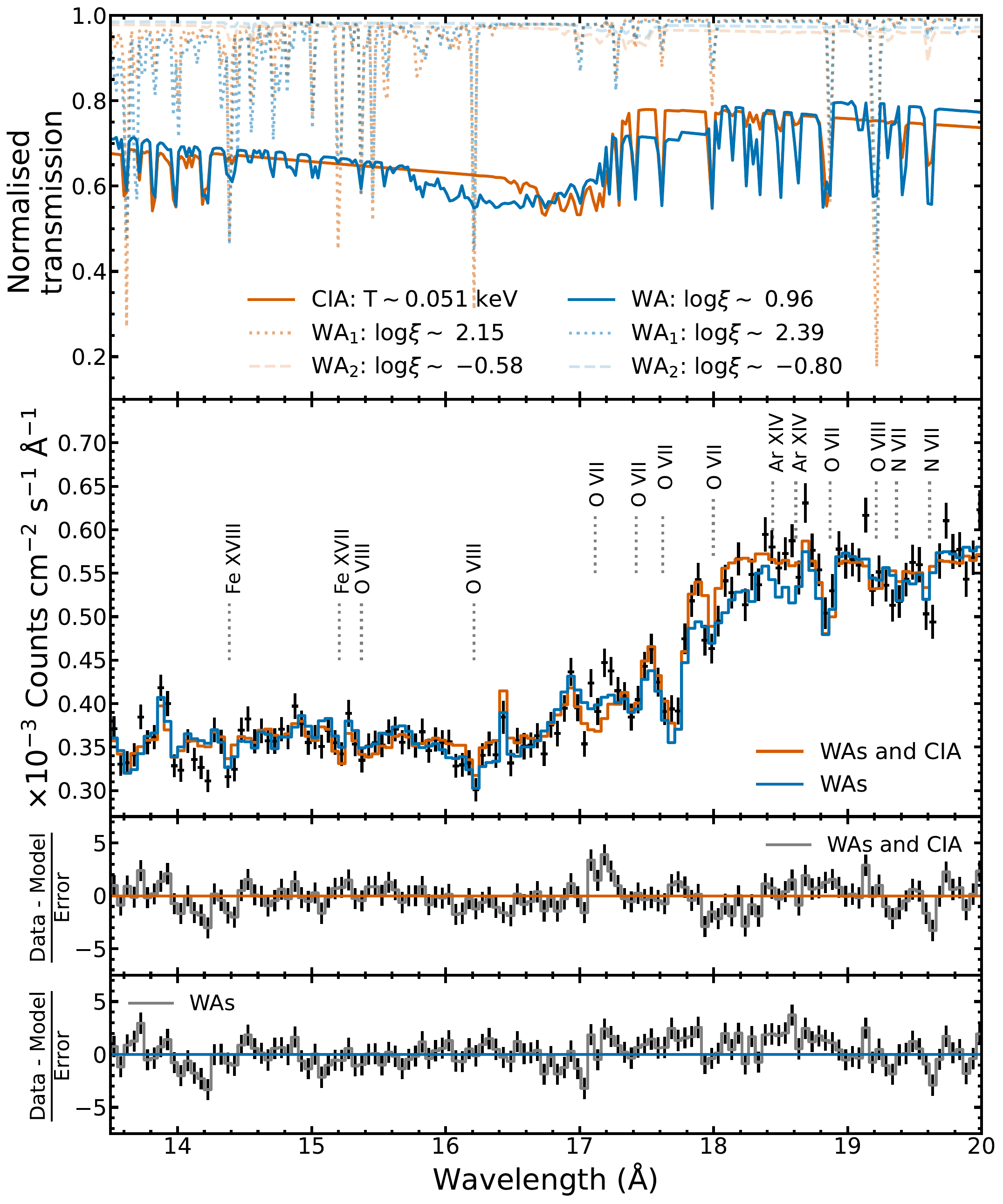}
    \caption{\textbf{Upper panel}: Transmission plot, where transmission is defined as the observed spectrum divided by the continuum. This plot compares how the CIA (red) behaves around the $13-20$ \AA\ region in comparison to a WA with $\log\xi\sim0.96$ (blue). Showcased in fainter red and blue are the transmissions obtained by the other WAs used in both models.  \textbf{Upper middle panel}: Difference between a model using a combination of WAs and CIA (red) as opposed to only WAs (blue). ``WAs and CIA'' is used to indicate our best-fit model. ``WAs'' is meant to represent the model with three WAs and no CIA. It is apparent that the latter case creates additional absorption in the $17-18$ \AA\ region. The most prominent absorption features identified by our models are marked and labelled. \textbf{Lower middle panel}: Residuals of the CIA and WAs model. \textbf{Lower panel}: Residuals of the WAs model. }
   \label{hotvsxabs}
\end{figure}

To decide between the models we calculated the $\Delta$AIC scores. The $\Delta$AIC represents the difference between the best model obtained, which is the one with the smallest Akaike criterion value (AIC$_\text{min}$), and each of the other models \citep{Burnham_2002}. The formula is given by $\Delta $AIC= AIC$_{i}$$-$AIC$_{\text{min}}$, where AIC$_\text{min}$ represents the smallest AIC value obtained, AIC$_i$ stands for the other AIC values found, and where all models must have the same degrees of freedom \citep{Akaike, Burnham_2002}. The larger the $\Delta $AIC, the less likely it is that two models might be equally as acceptable for the data. In fact, if $\Delta $AIC $> 10$, then we can assume that AIC$_i$ is not suitable when compared to AIC$_\text{min}$. Doing so, we concluded that a model including a CIA was more suitable as the obtained $\Delta$AIC$ = 82 \gg 10$ (see Table \ref{CIAWA}). Fig. \ref{hotvsxabs} provides visual representation of how a model without this additional collisionally ionised gas leads to additional absorption in the $17-19$ \AA\ region and how the addition of the CIA component results in a better fit and better residuals. Notice how the models shown in Fig. \ref{hotvsxabs} already include emission features (Sect. \ref{emission}). Table \ref{table:4} provides a summary of the best-fit parameter obtained for the absorption and emission components used in our model.

\begin{table}
      \caption{\label{CIAWA}Akaike test parameters for the WAs and CIA fitting. } 
\centering                                      

\centering                                   

\begin{tabular}{c c c c c c}         
\hline\hline   
        
Model & $C$-value  & d.o.f & $k$& AIC & $\Delta$AIC \\

\hline

Only WAs & $1325$   & $551$ & $21$&$1367$&$82$\\

WAs and CIA &$1243$ & $551$ & $22$&$1285$& 0 (AIC$_{\text{min}}$)\\

            \hline
\end{tabular} 
\end{table}

\begin{table}
\caption{\label{table:4} Best-fit free parameters for the RGS data fitting.}
\centering
\begin{tabular}{c c c}
\hline\hline

Component& Parameter & Value  \\

\hline

{\sc hot}\small{$_{\text{MW-cold}}$} & $N_\text{H}$ [cm$^{-2}$]& $ (1.98 \pm 0.05)\times 10^{20}$  \\
                                    &  $T$ [keV] & $ 1.0\times 10^{-6}$ [fixed] \\

\hline

{\sc pow} & norm & $(3.27 \pm  0.02) \times 10^{51}$ \\
           & [photons s$^{-1}$ keV$^{-1}$] & \\
           & $\Gamma_1$&$ 2.80 \pm  0.01$\\
           & $\Gamma_2$ & $2.51 \pm 0.05$\\
           & $E_{\text{break}}$ [keV]& $0.99 \pm 0.03$ \\

\hline

{\sc hot}\small{$_{\text{MW-warm}}$}& $N_\text{H}$ [cm$^{-2}$]& $ (1.0 \pm 0.3) \times 10^{19}$ \\
                                    & $T$ [keV] & $(5.00 \pm 1.40) \times 10^{-2}$\\
             
\hline

{\sc amol}\small{$_{\text{dust}}$}&  $N_\text{dust}$ [cm$^{-2}$]&  $(7.29 \pm 0.93)\times 10^{16}$ \\
                           &$v_{\text{out}}$ [km s$^{-1}$]& $270\pm110$ \\
 
\hline

{\sc xabs}\small{$_{\text{WA}_1}$}  &   $N_{\text{H}}$ [cm$^{-2}$]&  $(1.20 \pm 0.19) \times 10^{21}$ \\
                        & $\log\xi$ [erg cm s$^{-1}$] & $2.15 \pm 0.05$ \\
                        &  $C_V$\tablefootmark{a} &   $0.85\pm 0.09$  \\
                        & $v_{\text{out}}$ [km s$^{-1}$]&    $60 \pm 40 $ \\
\hline

{\sc xabs}\small{$_{\text{WA}_2}$}  &   $N_{\text{H}}$ [cm$^{-2}$] &  $(1.64 \pm 0.56 )\times 10^{20}$ \\
                    &  $\log\xi$ [erg cm s$^{-1}$]&  $-0.58$ $\pm 0.11$ \\
                    &  $C_V$\tablefootmark{a}  &   $\leq1$  \\
                    &  $v_{\text{out}}$ [km s$^{-1}$] &  $\leq 110$ \\

\hline

{\sc hot}\small{$_{\text{CIA}}$} & $N_{\text{H}}$ [cm$^{-2}$]& $(8.17 \pm 0.32)\times 10^{21}$ \\
                                &  $T$ [keV] & $ (5.07 \pm 0.31)\times 10^{-2}$  \\
                                &  $C_V$\tablefootmark{a} & $ 0.47 \pm 0.08$  \\
                                & $v_{\text{out}}$ [km s$^{-1}$]&   $ 250 \pm 70$  \\

\hline

{\sc pion}\small{$_{\text{em}}$} & $N_{\text{H}}$ [cm$^{-2}$] &  ($9.00 \pm 0.61) \times 10^{22}$ \\
            & $\log\xi$ [erg cm s$^{-1}$]&  $ 1.56 \pm 0.05$  \\
            & $\Omega/4\pi$\tablefootmark{b}  &  $ (3.84 \pm 1.31) \times 10^{-3}$   \\
            & $v_{\text{out}}$ [km s$^{-1}$] &  $\leq 40$  \\
\hline
\end{tabular}
\tablefoot{
\tablefoottext{a}{Unit-less parameter representing the covering factor of the absorber.}
\tablefoottext{b}{Aperture angle for emission. Although called $\Omega$ in \rm{SPEX}, it is meant to represent $\Omega/4\pi $}.}
\end{table}


\subsection{Emission lines}\label{emission}

The best-fit absorption model showed some residuals in emission, especially at the energies corresponding to O and N, which may indicate the presence of an emitting plasma, as can be seen from the middle panel of Fig. \ref{dust}. Therefore, we introduced a physically motivated emission model, {\sc pion}, which is able to calculate the transmission and emission of a slab of photoionised plasma where the ionic column densities are related through a photoionisation model \citep{Miller_2015, Mehdipour_2016}. Since the origin of the emission embedded in the absorption features of AGN is in general not known, we investigated two different physical scenarios for emission to test the connection between WAs and emission.

The first scenario, \textit{Case 1}, assumed that the emission is linked to the WAs and created in ionisation cones where re-emission occurs, resulting in narrow emission lines \citep{Behar_2003}. The geometry of this scenario would involve the emission being created within the ionisation cone of the WA, which is in the line-of-sight of the observer. Moreover, in this scenario the ionisation cone would be near the dusty torus, meaning the emission would not only be absorbed by the WA itself, but also by dust, as well as the ionised gas from the host galaxy and our own galaxy.

The second scenario, \textit{Case 2}, is one where emission is created in the narrow line region (NLR). However, this time emission would not be absorbed by the galaxy material as it would originate from within some photoionised plasma situated far away from the ionising source. Therefore, it would not be absorbed by any dust or gas in the system.

We investigated these two different scenarios by creating models where we related the {\sc pion} component differently with the other components in other to replicate the emission cases above. To evaluate which emission case was the most appropriate, we used once more the Akaike criterion. We found that both scenarios were equally as likely to be the reason for the emission being present in the spectra (Table \ref{Pion emission c-stats}). As discussed later in Sect. \ref{Discussion}, we adopted \textit{Case 2} while modelling the spectrum, which improved our $C$-stat/d.o.f. from $1243/551$ for the model without emission, to $1000/547$ for the model accounting for emission. The addition of this emission also altered the parameters of some of the components in the model, to be those shown in Table \ref{table:4}.

\begin{table}
      \caption{\label{Pion emission c-stats}\, Akaike test parameters for the {\sc pion} fitting.} 
\centering                                      

\centering                                   

\begin{tabular}{c c c c c c}         
\hline\hline   
      
Model & $C$-value  & d.o.f & $k$& AIC & $\Delta$AIC \\
           
\hline
            
\textit{Case 1} &$995$  & $547$ & $25$&$1045$&$0$ (AIC$_{\text{min}})$ \\
            
\textit{Case 2} &$1000$ & $547$ & $25$&$1050$& 5\\

\hline
        \end{tabular} 
\end{table}


%

\begin{figure}
 \centering

	\includegraphics[angle=0,width=1.0\hsize]{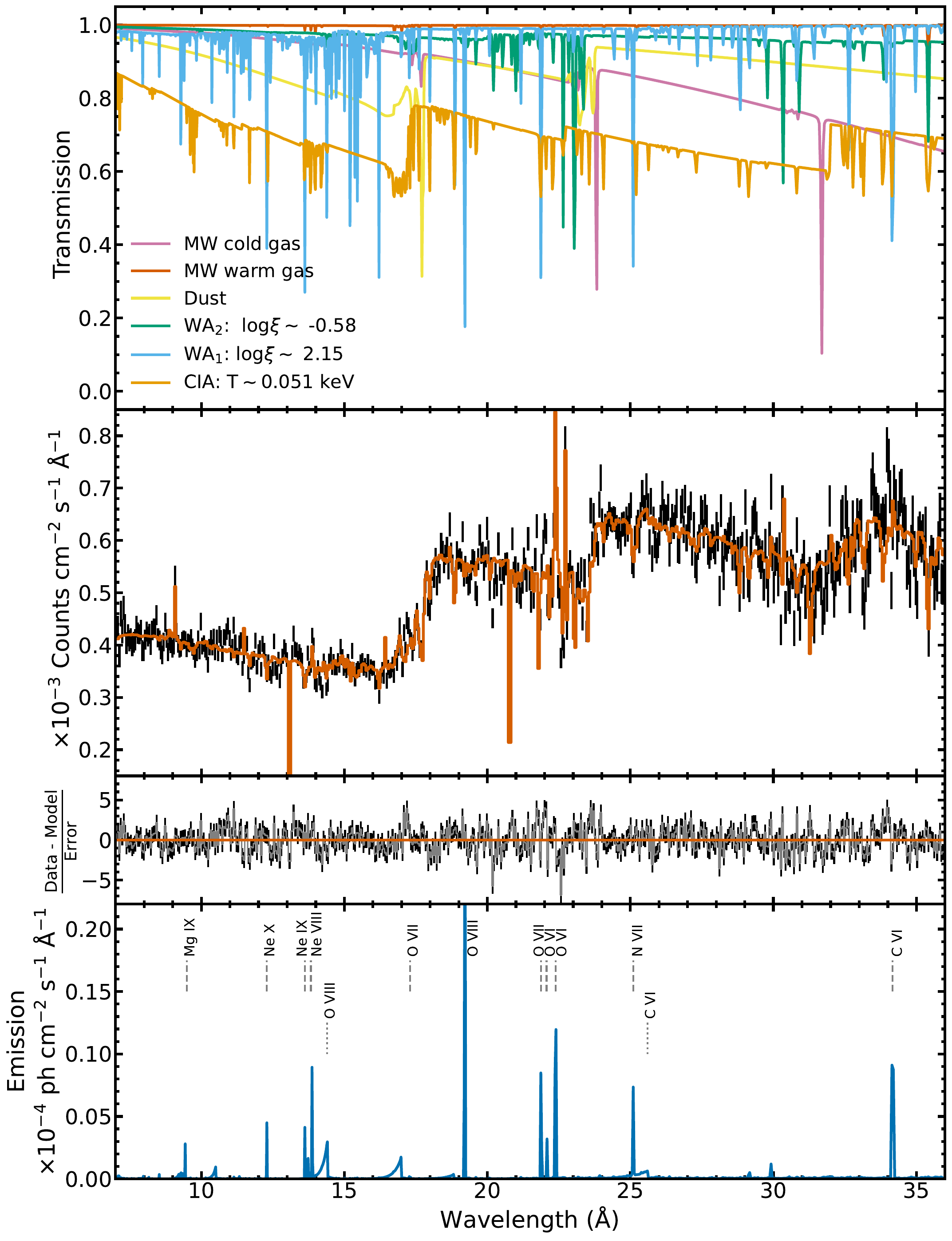}
    \caption{\textbf{Upper panel}: Transmission plot, where transmission is defined as the observed spectrum divided by the continuum. The different absorbing components are explained in the legend. \textbf{Middle upper panel}: The RGS data (black) with the best-fit model (red). \textbf{Middle lower panel}: The residuals from our best-fit model. \textbf{Lower panel}: Emission features predicted by {\sc pion} in emission, with the most prominent ones being labelled.}
   \label{Transmission}
\end{figure}

In Fig. \ref{Transmission}, we show the transmission of all the best-fit absorbing components used in our model together with the data and the emission features that were predicted to be in the data when using {\sc pion} in \textit{Case 2}. Table \ref{table:4} contains the parameters obtained for our best fit. 

\section{Discussion}\label{Discussion}

We have presented an analysis of the OM, EPIC-pn, and RGS spectra for the narrow-line Seyfert 1 AGN Mrk\,766. The high spectral resolution of the RGS gratings allowed us to find a plethora of absorption features, originating both from the Galaxy and intrinsic to Mrk\,766. 

The intrinsic absorption affecting the soft X-ray spectrum of Mrk\,766 comprises two phases of WAs, dust, and a warm CIA. We noted that none of the WAs seem to be outflowing at large speeds, and that their column densities and ionisation parameters are similar to the commonly observed ranges in Seyfert 1 AGN \citep{Mckernan_2007, Laha_2014}. Moreover, in order to reproduce the intricate shape of the spectrum, we allowed for partial covering for all of the absorbers. Finally, we investigated the presence of a photoionised emitting gas with $\log\xi= 1.56\pm0.05$, that is responsible for the emission features present in the spectrum. 

For the analysis, we used a time-averaged spectrum consisting of observations with negligible changes in the continuum spectral shape. The broadband flux did, however, somewhat vary between the individual observations.

\subsection{The complex absorbers in Mrk\,766}

\subsubsection{Dust}

The soft X-ray spectrum of Mrk\,766 has been subject of debate due to the superposition of several components, mimicking a very structured spectrum, e.g. around $17$, $25$ and $30$ \AA. Although initial studies associated this intricate structure to WAs \citep{George_1998}, \cite{Raymont_2001} used RGS data to argue that the spectra of this source should be explained by strong, highly relativistically broadened Ly $\alpha$ emission lines of \ion{H}-like \ion{O}, \ion{N}, and \ion{C}\ from the near vicinity of a Kerr black hole. This scenario was corroborated  by \cite{Sako_2003}, who also reported that a relativistic line model successfully reproduced the spectra of Mrk\,766. \cite{Raymont_2001} obtained a similar conclusion for the source MCG--6-30-15, which showed a similar spectrum in the RGS band.

On the other hand, \cite{Lee_2001} studied the \textit{Chandra}-HETG spectra of MCG--6-30-15 and argued that a dusty warm absorber model provided a comprehensive explanation for the spectral shape of this source. \cite{Lee_2001} concluded that MCG--6-30-15 contained dust that was embedded in a partially ionised WA, and believed that it could be possible that Mrk\,766 hosted a similar system. In our model we were able to identify the presence of both, dust and WAs. Both of these components are necessary to account for the structured spectrum of Mrk\,766, in particular the absorption features in the $17$ \AA\ region.

We find the molecular column density of the dust to be $N_{\rm dust}=(7.29\pm0.93)\times 10^{16}$ cm$^{-2}$. Nevertheless, it is not straightforward to assign this dust component to the host Galaxy. In a Milky Way-type environment, a dust component would be naturally associated with cold gas. Then, the observed relation between $E(B-V)$ and $N_{\text{H}}$ in our Galaxy could provide us with this gas' component column density \citep{Bohlin_1978, Costantini_2022}. However, for this source, an SMC extinction law \citep{Gordon_2003} seems to better describe the OM data (Sect. \ref{Sect 3}), making thr relation between $E(B-V)$ and $N_{\text{H}}$ difficult to determine.

A direct fit to the RGS data using an additional cold gas component intrinsic to the source resulted in a shallow absorption ($\sim1\times10^{20}$ cm$^{-2}$). The comparison between the dust and the extra gas' column density lead to an unrealistic overabundance of iron ($\times 40$), against previous estimates of $\approx(1.5-15)$ solar for spiral galaxies \citep{Grasha_2022_2022ApJ...929..118G}. Moreover, the cold gas' linked \ion{O}{I} absorption line is also not visible in the data. The association of our dust component with a Galactic-type gas and dust mixture is therefore not likely. This could indicate that the dust we see is not associated with the host galaxy, but rather to more nuclear regions of the system. Interestingly, we find the dust to be outflowing with a velocity of $270\pm110$ km s$^{-1}$. This velocity is of the same order of magnitude as that of the CIA, suggesting a potential association between these two components.

Historically, Mrk\,$766$ has exhibited filaments, wisps, and irregular dust lanes around an unresolved nucleus \citep{Malkan_1998, Riffel_2006}, along with extended optical and radio emission \citep{Ulvestad_1995, Gonzalez_1996, Nagar_1999}. Additionally, in the near-infrared band (NIR), its continuum emission displays a complex shape, influenced by contributions from the central engine as well as circumnuclear stellar populations and dust \citep{Rodriguex_2005, Riffel_2006}. Therefore, it is also possible that this dust is linked to other complex environments of the source.

\subsubsection{Collisionally ionised absorber}

Looking at the breakdown of the models in Fig. \ref{Transmission}, one can clearly see that the most prominent absorber appears to be the collisionally ionised one (orange line in the upper panel of Fig \ref{Transmission}). The inclusion of the CIA significantly improved the over-absorption around $18.5$ \AA, although from the residuals in Fig. \ref{hotvsxabs}, one could argue that it also worsened the fit at $17.2$ \AA, where both models seem to miss a spectral feature.

Although this would be the first time that a CIA is detected in Mrk\,766, its presence is not unforeseen, as other studies have already detected absorbers of this sort in other AGN. \cite{Ogorzalek_2022} found 6 different absorbers intrinsic to NGC 4051, two of them CIAs. Their collisionally ionised absorbers appeared to be outflowing at velocities in the same order of magnitude as the one found for our CIA, and had a similar temperature as the one found in this work. Previous to this, signatures of post-shock gas cooling were detected for NGC 4051 \citep{Pounda_2011, Pounds_2013}, a conclusion that led  \cite{Pounds_2013} to postulate that such signatures could be due to fast, highly ionised winds, probably created in the vicinity of the supermassive black hole, that lost their mechanical energy after shocking against the ISM at a small enough radius for strong Compton cooling. But NGC 4051 is not the only Seyfert with collisionally ionised gas present in the system, other sources have also been found to contain similar CIA shocked gas, including the radio-loud NGC 4151 \citep{Wang_2011}, Mrk 573 \citep{Paggi_2012}, or NGC 3393 \citep{Maksym_2019}.

The origin of this CIA is not possible to estimate with our current model, although we are able to determine its outflowing velocity to be $250\pm70$ km s$^{-1}$, and therefore we can place constraints on its distance. Using the methodology typically applied to WAs as outlined by \cite{Blustin_2005}, we can approximate the lower limit of the distance for CIAs. This is feasible because the necessary parameters for this calculation are known and available from our model. Therefore, assuming that the outflowing velocity ($v_{\text{out}}$) is larger than or equal to the escape velocity \citep{Blustin_2005},  we found this CIA to be positioned somewhere beyond $0.11$ pc. Although it is not clear where this absorber originates, it could be possible that it arises from a shocked interface between a WA and the host galaxy \citep{King_2015}. It could also be the case that the CIA is created during a shock between the gas and the IGM \citep{Ogorzalek_2022}. We find the column density of this CIA is measured to be $N_{\text{H}}= (8.17\pm0.32)\times 10^{21} \text{cm}^{-2}$, which is of the same order of magnitude as the other identified photoionised absorbers. We find its temperature to be $T= 50.7\pm3.1$ eV and to be partially covering $(47\pm8)\%$ of the source, indicating how it is possible for this absorber to have a patchy / stratified structure.

\subsubsection{Warm absorbers and comparison with previous studies}

X-ray analysis carried out in the past pointed out the presence of two or three warm absorber components, albeit with slightly differing parameters. We note however that past works depend on different instrument characteristics and overall modelling. None of the previous analyses considered using a collisionally ionised absorber alongside the warm absorbers and only some implemented partial covering to their models, as well as dust. Moreover, each study carried out their own, and therefore different, SED modelling, which in turn strongly affects the ionisation parameters obtained during fitting.

Historically, at least two phases of WAs have already been detected in the soft X-ray spectrum of Mrk\,766. For example, \cite{Laha_2014} analyzed \rm{XMM}-\textit{Newton} data from 2001. They found three phases of warm absorbers at $\log\xi\sim -0.70$, $\log\xi\sim -0.94 $, and $\log\xi\sim 1.35$. These absorbers seemed to be outflowing with velocities of $v_{\text{out}}\sim 810$ km s$^{-1}$, $v_{\text{out}}\sim 1020$ km s$^{-1}$, and $v_{\text{out}}\sim 0$ km s$^{-1}$ respectively. 

Furthermore, \cite{Winter_2012} used \textit{Suzaku} data from $2006$ and reported the presence of two warm absorbers at $\log\xi\sim 2.85 $, and $\log\xi\sim 1.95$, while \cite{Mckernan_2007} analysed $2001$ \textit{Chandra} observations to find the presence of three WAs at $\log\xi\leq 0.6 $, $\log\xi\sim 2$, and $\log\xi\sim 3.1$ with outflowing velocities consistent with zero. Mrk\,766 is known for being a variable AGN, meaning that its flux fluctuates between periods of higher and lower fluxes \citep{Leighly_1996}. This could in turn affect the properties of the WAs present in the source, making them vary in the ionisation parameters. However, previous studies have found that in other sources, changes in X-ray flux do not necessarily correlate with WAs variability \citep{Kaspi_2004,Constantini_2007,Longinotti_2013, Silva_2018}. \cite{Buisson_2018} also investigated the X-ray spectrum of Mrk\,766 with data from \rm{XMM}-\textit{Newton}, \textit{Swift}, and \textit{NuSTAR} to find that a hybrid model with a combination of two phases of WAs with partial covering provided a good representation for the absorption present in the spectrum.

\begin{figure}
	\includegraphics[scale=0.22]{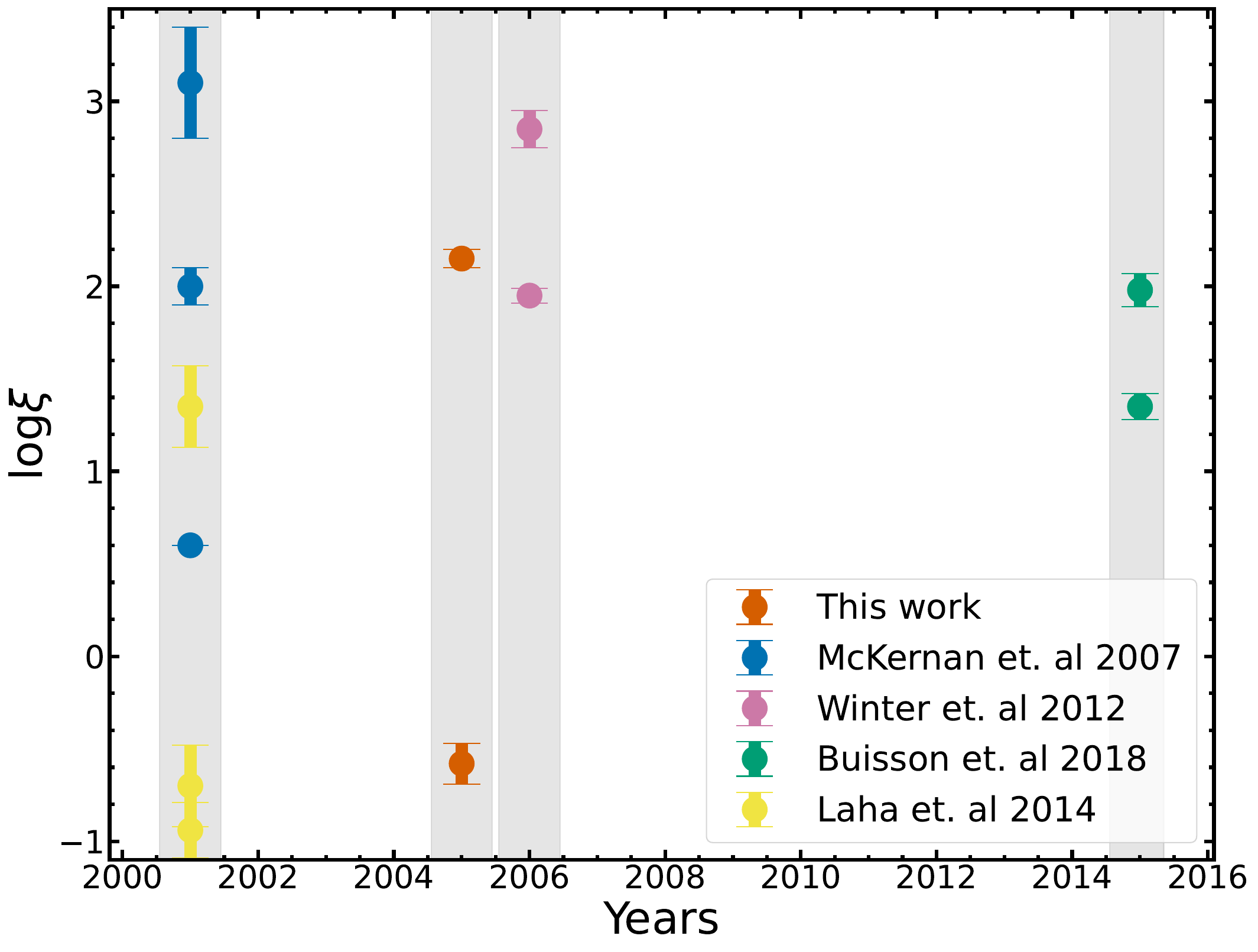}
    \caption{Ionisation parameters of the  warm absorbers identified in Mrk\,766 through the years. The results from this work are shown in red. We can see that the higher ionised absorber found in this work seems to be in agreement with other WAs found in the literature, while our less ionised WA does not correlate to others previously found.}
   \label{comparison}
\end{figure}

Fig. \ref{comparison} places the WAs findings of this work with respect to those found in the literature. Despite the differences in results, it appears that in the different energy bands covered by the different instruments, there seems to exist a long lived absorber with $\log\xi\sim 2$.

\subsubsection{Warm absorbers' distance estimation}

The location of WAs in AGN is in general unconstrained. However, taking various assumptions regarding the morphology and other aspects of these X-ray absorbers, it is also possible to put constraints on the lower and upper limit where individual WAs may lie in the system \citep{Blustin_2005}.

The first assumption that is required is that the depth of the X-ray absorber, $\Delta r$, should not exceed the distance at which the absorber lies from the source \citep{Reynolds_1995}. This way, taking the hydrogen column density to be $N_\text{H}=n\Delta r f$, where $n$ represents the column density of the gas, and $f$ the filling factor, one can follow the method outlined in \cite{Blustin_2005} and use the parameters of our best-fit model alongside the ionising luminosity obtained from the EPIC-pn and OM data ($\log L_\text{ion}=44.20$), to obtain the upper limits for the distances where the WAs may be found.

Therefore, the upper limit for the WA lying at $\log\xi$$ = 2.15\pm0.05$ was found to be at a distance from the ionising source of $r\leq 395$ pc, while the WA at a $\log\xi$$=-0.58\pm0.11 $ obtained a looser constraint with an upper limit of $r\leq 2\times 10^6$ pc. This latter constraint would place this WA outside the host galaxy boundaries. However, it is possible that this outflow could be a galactic-scale outflow, which is seen as a warm absorber in the X-ray band, such as those in 1H 0419-577 \citep{Gesu_2017}.

To be able to calculate the lower limit of the position of the absorbers, we must assume that the outflowing velocity ($v_{\text{out}}$) is larger than or equal to the escape velocity. For WA1 and WA2 we adopted the upper limits on $v_{\text{out}}$ (Table \ref{table:4}). Adopting a black hole mass of $1.26\times10^6 M_{\odot}$ \citep{Giacche_2015} we obtain $r\geq 0.90$ pc and $r\geq1.08$ pc for the $\log\xi$$=-0.58\pm0.11 $ and $\log\xi$$ = 2.15\pm0.05$ absorbers respectively.

\subsection{Emission lines}

It is not possible to distinguish, on a statistical basis, if the emitting gas is part of the absorbing gas or if it is a separate, unabsorbed component. However, comparing the values of the individual {\sc pion} parameters with those of the WAs, it became apparent that the $\log\xi$ of the emission plasma was not consistent, within more than $3\sigma$ with neither of the ionization parameter of WA$_1$ and WA$_2$. This in itself indicated that this emission might not be linked to the WAs, making \textit{Case 1} less likely (see Sect. \ref{emission}). Moreover, for both scenarios we do not expect the emission lines to display a significant outflow velocity. We found this component to have a $\log\xi= 1.56\pm0.05$ and $N_\text{H}=(9.00\pm0.61)\times 10^{22}$ cm$^{-2}$, as well as an outflowing velocity consistent with zero, and an aperture angle $\Omega/4\pi=(3.84\pm1.31) \times 10^{-3}$, which is consistent with the idea that the emission detected was produced in a narrow cone within the NLR. Using {\sc pion} allowed us to detect emission from H and He-like transitions of N, O, C, and Ne as well as the RRCs of \ion{O}{vii}, and \ion{O}{viii}. The detection of these narrow RRCs imply that the gas creating this emission must be photoionised, and that, therefore, an association of the emitting gas with the absorbing CIA is unlikely. This is because RRCs tend to be broad features for hot, collisionally ionised plasmas but are narrow features for cooler photoionised plasmas \citep{Liedahl_1996, Liedahl_1999}. Moreover, an independent fit of the narrow RRC features ({\sc rrc} model in SPEX) hints that they originated at a low temperature ($6.0\pm0.2$ eV). This temperature is associated with a photoionised gas, which is consistent with the {\sc pion} modelling. Additionally, this would disfavour an association with the CIA component as, in that case, we would have expected broadened RRC features.  

\begin{figure}
	\includegraphics[scale=0.63]{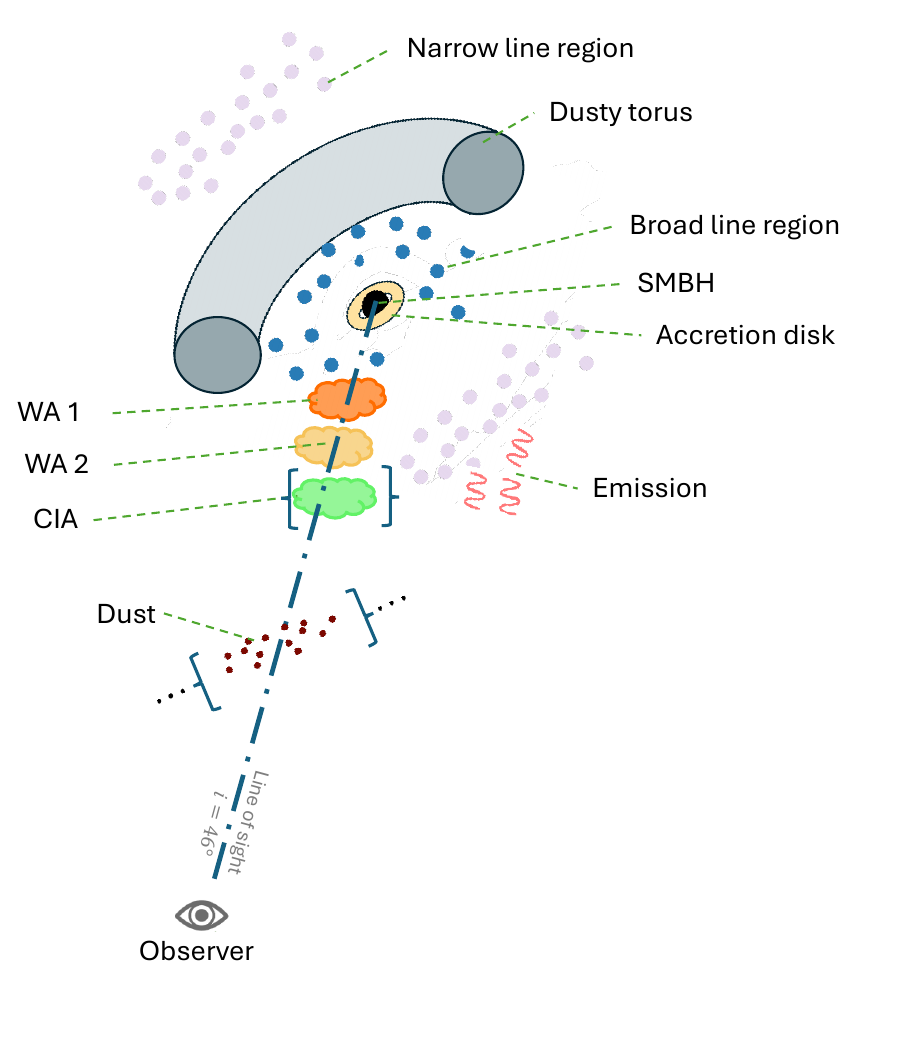}
    \caption{Depiction of the emission for \textit{Case 2}, where the emission is not linked to the WAs and is created away from our line-of-sight. Therefore, it is not absorbed by the components of the source and host galaxy. The CIA is placed within brackets to showcase that this position is only a placeholder, as from our analysis we could not constrain its position with respect to the rest of the system. The key parts of the AGN are clearly labelled, while the different absorbing components are given different colours and patterns to differentiate them.}
   \label{system}
\end{figure}

The tight degeneracy that exists between the distance from the ionising source and the gas density implies that, if we assume that this gas is located in the NLR ($100$ pc), then, given its ionization parameter, the density must be small ($1\times10^{3}$ cm$^{-3}$) in order to have such a high ionising parameter. It could also be the case that the emitting gas is far from our line of sight, but could be closer to the black hole than the WA. A representation of this scenario can be seen in Fig. \ref{system}, where all physical absorbing and emitting components that make up the model are illustrated.

\section{Conclusions}

In this paper we have performed an analysis of the high energy resolution X-ray spectrum of the narrow-line Seyfert 1 galaxy Mrk\,766 using 4 observations taken with \rm{XMM}-\textit{Newton} in 2005, to investigate the properties of the complex ionised absorber / emitter along the line of sight as well as absorption by dust intrinsic to the source. From our findings we conclude the following:

\begin{itemize}

\item There are two distinct phases of WAs present in Mrk\,766, with $\log\xi$ $= 2.15\pm0.05$ and $\log\xi$ $=-0.58\pm0.11$. They have moderate outflowing velocities with $v_{\text{out}}=60\pm40$ and $v_{\text{out}}\leq 110$ km s$^{-1}$ respectively and they both partially cover the source with $C_V = 0.85\pm0.09$ and $C_V \leq 1$, indicating potential stratification in their structure. \\

\item We find evidence for a collisionally ionised absorber present in Mrk\,766 with $T= 50.7\pm3.2$ eV and $N_{\text{H}}= (8.17\pm0.31)\times 10^{21}$ cm$^{-2}$. We find its velocity to be of $v_{\text{out}}= 250\pm70$ km s$^{-1}$. This could be created from a shocked interface between a WA and the host galax, or from ionised gas that was shocked against the ISM. \\

\item There is a dust component present in Mrk\,766. It is possible that this dust may not be associated with the host galaxy, but rather to more nuclear regions of the system.
 \\

\item We find that emission is present in the spectrum of Mrk\,766 and that it is likely created by a photoionised gas with $\log\xi= 1.56\pm0.05$ in the NLR.\\

\end{itemize}

\begin{acknowledgements}
     We thank the anonymous referee for their thorough reading and for their very useful comments and suggestions that have improved the clarity of the paper.
\end{acknowledgements}

%
%

\bibliographystyle{aa} 
\bibliography{MRK766}

\end{document}